# Post-transcriptional Regulation Drives Temporal Compartmentalization of the Yeast Metabolic Cycle

Maria Concetta Palumbo[1], Lorenzo Farina[2,¶], Alberto De Santis[2], Alessandro Giuliani[3], Alfredo Colosimo[1], Giorgio Morelli[4] and Ida Ruberti[5]

[1]Department of Physiology and Pharmacology, Sapienza University of Rome, Rome, Italy.
[2]Department of Computer and Systems Science "Antonio Ruberti", Sapienza University of Rome, Rome, Italy. [3] National Institute of Health (ISS), Department of Environment and Primary Prevention, Rome, Italy. [4]National Research Institute for Food and Human Nutrition, Rome, Italy. [5]National Research Council, Institute of Molecular Biology and Pathology, Rome, Italy.

[¶] Corresponding author: *Lorenzo Farina, Dipartimento di Informatica e Sistemistica "Antonio Ruberti", Via Eudossiana 18, 00184 Roma (Italy). Tel: +39-06-4457526 Fax: +39-06-44585367. E-mail:* lorenzo.farina@uniroma1.it

**Abstract**

The maintainance of a stable periodicity during the yeast metabolic cycle involving approximately half of the genome requires a very strict and efficient control of gene expression. For this reason, the metabolic cycle is a very good candidate for testing the role of a class of post-transcriptional regulators, the so called PUF-family, whose genome-wide mRNA binding specificity was recently experimentally assessed. Here we show that an integrated computational analysis of gene expression time series during the metabolic cycle and the mRNA binding specificity of PUF-family proteins allow for a clear demonstration of the very specific role exerted by selective post-transcriptional mRNA degradation in yeast metabolic cycle global regulation.

**Introduction**

Self-sustained temporal structured activities displaying a clear and robust periodic time behaviour, are perhaps the most time-honored objects in science. In fact, the dawn of modern science is generally assumed to coincide with the development of general laws able to explain and predict the regular orbits of planets. On a more empirical ground, the accomplishment of a task with a high degree of accuracy and robustness requires the presence of control mechanisms and strategies for actively sustaining the desired time profiles (*i.e.* a trajectory tracking device) such as regular oscillations, in the presence of variable and largely unpredictable environmental stimuli and noise.

*Saccharomyces cerevisiae* (yeast) metabolic cycles can be considered as very reliable temporal structures in which an extremely complex apparatus – the yeast cell – strives to maintain self-sustained cycles at different functional levels, ranging from gene expression (recently assessed on a genome-wide scale by the McKnight group [1]) to metabolic activity (oxygen consumption). The accuracy and robustness of such cycles, in terms of frequency and amplitude, is much greater than any other known cycle at cellular level, namely much greater than the reproductive cell cycle. For this reason, gene expression dynamics of the yeast metabolic cycle is a natural candidate for studying the "basic principles" of gene expression regulation. As a general rule, there is often the need of combining both positive and negative control actions to keep a stable and reliable trajectory in time. For example, in order to achieve fast, accurate and reliable vehicle dynamics, an integrated throttle and brake control system is usually required. This is the case, for example, of high performance race driving techniques – like *heel-and-toe* or *left-foot-braking* – which rely on simultaneous and/or alternate gas/brake pedal usage. Clearly, such driving style is energy-consuming, so that it is worth using only when precisely regulated levels are required in the face of sudden and unpredictable external events. In other words, a *fast, precise and robust* behavior is very expensive both in terms of control strategies complexity and resource consumption.

When dealing with gene expression data, the "throttle pedal" may correspond to the transcription factor (TF) system, allowing for specific DNA sequences to be made accessible to polymerases and synthesizing the corresponding mRNA, whereas the "brake pedal" may be associated with the so called degradosome system, made up of those factors able to selectively degrade specific mRNA molecules. Indeed, there is a growing interest in the study of post-transcriptional regulation [2] and especially for the mechanisms leading to the modulation of mRNA turnover in response to environmental changes [3,4].

Studies on the regulation of mRNA stability during the yeast reproductive cell-cycle have been carried out on some specific transcripts, such as the histone mRNAs [5]. On a global scale, a recent computational analysis [6] showed a specific and active role for transcript stability regulation in the yeast reproductive cell cycle. Evidence of post-transcriptional regulation of the histone genes has also been shown in higher eukaryotes and mammalian cell types, including HeLa cells [7] and *Drosophila* circadian rhytms [8]. However, experimental evidence of global, cell-cycle specific, post-transcriptional regulation of gene expression is still lacking.

Recently, five member of the RNA binding proteins PUF family (PUF1-PUF5), or Pumilio-homology domain, has been studied for sequence specificity on a genome wide-scale [9] thus providing insight into one of the most important mechanism of post-transcriptional regulation. The number of the identified targets are 40 for PUF1, 146 for PUF2 and around 200 for PUF3, PUF4 and PUF5 indicating that the expression of a large number of genes can be, in principle, modulated by specific post-transcriptional events. PUF proteins, as mRNA-specific regulators of deadenylation, has been conserved throughout eukaryotes suggesting that they are likely to play a prominent role in the control of transcript-specific rates of deadenylation in yeast by interacting with the mRNA turnover machinery [10]. A recent integrated analysis performed by De Lichtenberg et al. [11] on yeast cell reproductive cycle data, did not show any relevant correlation of PUF family genes with the cell cycle. In fact, none of the PUF1-5 genes are present in their extended list of 1159 reproductive cell-cycle regulated genes. Obviously, we do not exclude the possibility of a specific role for post-trascriptional regulation on a global scale, but no conclusions can be drawn basing upon current available data on PUF family targets.

Gene transcription has received the most attention for historical and technical reasons, but transcription is just the first stage in the process of gene expression. From splicing to polyadenylation, every aspect of a transcript's life is subject to elaborate control and it is therefore no surprise that many cellular factors and mechanisms are devoted entirely to modulating the rate of mRNA degradation [2]. Consequently, it is of paramount importance to fill this gap and provide evidences of the mode of action of the other "arm" of gene regulation during gene expression temporal programs. Given the biological importance of the presence of links and coordinated actions across multiple layers of control, understanding gene expression requires an integrated view by combining data from different aspects of regulation [12]. Although this approach holds great promise, there are currently few studies that take into account regulation at multiple levels [12].

Here we investigate the role of the "PUF control system" [9], in the temporally compartmentalized regulation of gene expression during the yeast metabolic cycle [1]. Our integrative computational analysis demonstrates a clear phase specificity of PUF-mediated regulation in the yeast metabolic cycle, allowing us to hypothesize post-transcriptional regulation as a key player in coordinating the global timing of gene expression during specific phases of the metabolic cycle.

**Materials and Methods**

The microarray gene expression data relative to the Tu *et al*. paper [1] were downloaded from the website http://yeast.swmed.edu/cgi-bin/dload.cgi. Following the authors indication, we selected their sentinel genes as probes for the three phases in which the metabolic cycle was partitioned; these genes were MRPL10 (RB phase), POX1 (RC phase) and RPL17B (OX phase), namely RB: reductive biosynthetic, RC: reductive charge, OX: oxidative. For each of the three sentinel genes we selected a cluster containing the 500 mostly correlated gene products so to obtain a 1500 genes data set. Each cluster belongs to one of the three phases of the metabolic cycle. The role of PUF genes during the metabolic cycle regulation was assessed by means of standard statistical tests, namely the correlation of PUF mRNA temporal variation with the centroid profile of the three clusters and the computation by means of the odds ratio (OR) statistics for the 'enrichment' of gene pairs sharing the same PUF within the same phase of the cycle. The binding specificity of PUF genes was assessed on the basis of Gerber *et al*. work [9]. The PUF based results were compared with similar analysis based on TFs. The binding specificity of TFs was based upon McIsaac's *et al*. work [13] by considering a stringent $P$ value for DNA binding less than 0.001.

**Results and Discussion**

For each of the 'sentinel genes' of the three phases of the cycle the 500 most correlated genes in terms of Pearson correlation coefficient along temporal profiles were selected so giving rise to three clusters. The temporal profile of each cluster is depicted in Fig.1A in terms of cluster centroid activation together with the standard deviation. Data are expressed as Z-scores, that is each gene expression value is de-meaned and divided by its standard deviation. The genes pertinent to each cluster are correlated with the sentinel gene with a Pearson correlation coefficient $R$ ranging from 0.82 to 1 (RB), 0.87 to 1 (RC), 0.85 to 1 (OX) so giving a reliable picture of the three phases in time. The data are averaged over the three available cycles.

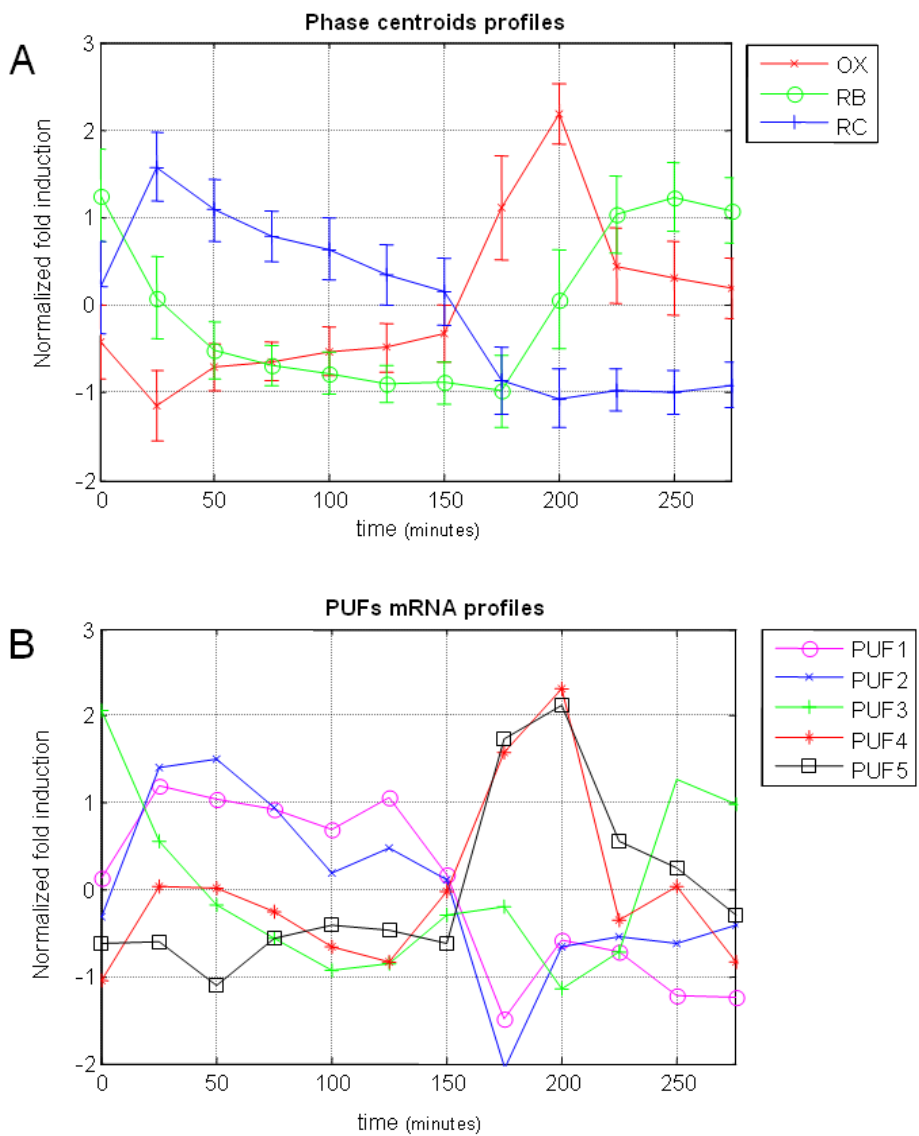

**Figure 1.** Temporal profiles of the metabolic phases centroids (A) and of the five members of the PUF family (B) as obtained by the same experimental dataset. Data have been Z-normalized (zero mean, unit standard deviation).

As it is apparent from Fig.1B there is a clear similarity between PUF time profiles and cluster centroids dynamics. Basically PUF1 and PUF2 covary with RC, PUF4 and PUF5 go together with OX while PUF3 covaries with RB phase.

Our statistical analysis provides empirical indication of temporal covariation between PUF family genes and yeast metabolic cycle. In order to go more into deep, we need to discriminate between a pure episodical and a potentially biological significant link, and therefore we move a step further by computing the Pearson correlation coefficient among all the gene pairs resulting from the three clusters corresponding to the three metabolic phases. By doing so, we ended up with more than one million (1124250) distinct correlation coefficients, that are distributed as in Figure 2.

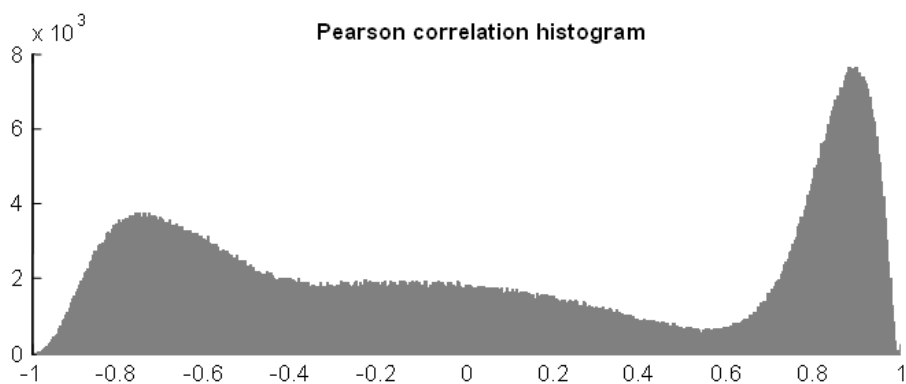

**Figure 2.** Pearson correlation histogram of all genes profiles pairs (1124250) obtained using the selected 1500 metabolic-cycle regulated genes.

As shown in Figure 2, there is a clear bimodal distribution allowing for an unambiguous partition of the gene pairs into three classes:

1) Positively correlated pairs ($r > 0.6$)
2) Linearly independent pairs ($0.6 > r > -0.6$)

3) Negatively correlated pairs (r < -0.6).

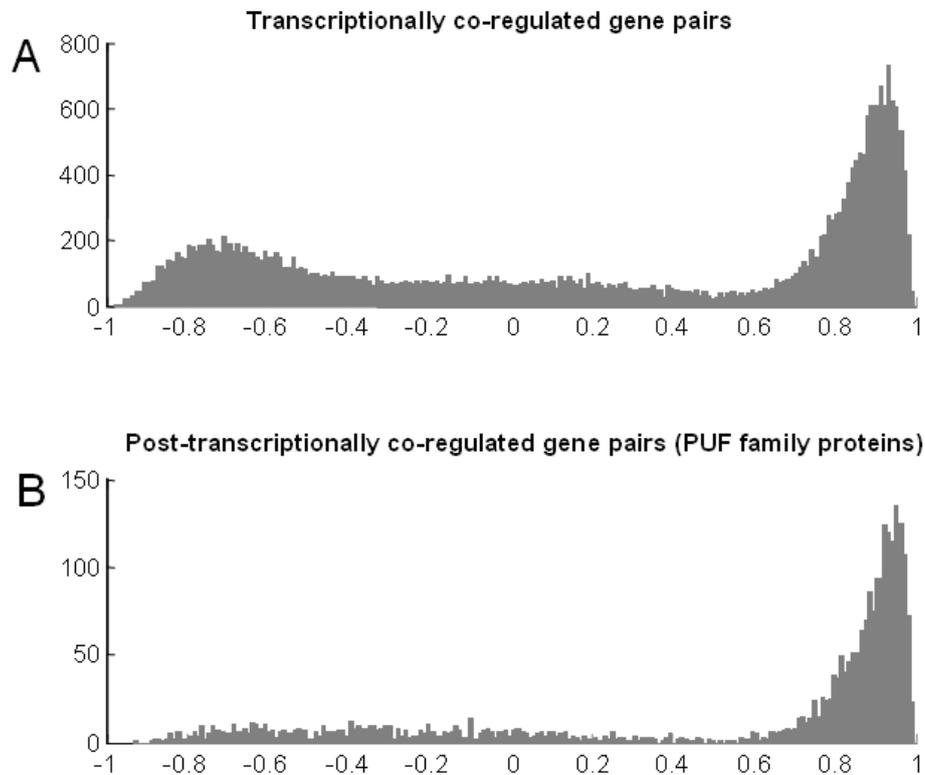

**Figure 3.** Pearson correlation histogram (A) of gene pairs sharing at least one transcription factor and (B) gene pairs sharing at least one mRNA binding protein of the PUF family (PUF1-5).

This partition comes directly from the existence of three temporal clusters, so that gene pairs relative to the same cluster go in class 1, while genes in a pair coming from different clusters, go alternatively in class 2 and class 3 depending on the relative phase shift of the corresponding clusters.

Such very sharp distribution together with the numerosity of the correlation coefficient data set allow us to consider two different filters on Figure 2 so to immediately highlight the role of post-transcriptional regulation as compared to TF based regulation. Consequently, from the initial population of Figure 2 we selected only those pairs sharing at least one TF in common (Figure 3A) and pairs sharing at least one PUF in common (Figure 3B).

As it is evident by the comparison between panels A and B of Figure 3, the pairs with a common PUF are by far more rich in the positively correlated class, while the pairs sharing the same TF are scattered along the distribution occupying all the three correlation classes. This results points to the highest specificity of PUF mediated post-transcriptional control with respect to TF control. Clearly, we must take into account the fact that we do have the binding specificity for much more TFs (118) than post-transcriptional regulators such as PUFs (5). In order to get an unbiased estimation of the different specificity of the two regulation systems, we computed the odds ratios for the TF and PUF commonality, respectively:

- OR (TF) = fraction of gene pairs having at least one common TF in the correlated subset w.r.t. all gene pairs having at least one common TF / fraction of pairs in the correlated subset w.r.t. all gene pairs.
- OR (PUF) = fraction of gene pairs having at least one common PUF in the correlated subset w.r.t. all gene pairs having at least one common PUF / fraction of pairs in the correlated subset w.r.t. all gene pairs.

A value equal to 1.37 for TF and 2.11 for PUF was obtaining, again pointing to a greater specificity of PUF control. This result was confirmed by the OR values computed for the anticorrelated pairs that was equal to 0.87 for TF and 0.28 for PUF.

The regulation specificity of PUF family members is mainly driven by PUF3 having an OR equal to 2.49 as for the enrichment of correlated pairs and 0.04 as for the depletion in anticorrelated subsets. It is also worth noting that PUF2 it is mildly enriched for anti-correlated pairs (OR =1.64) thus suggesting a regulatory role between OX and RC phases, and such role is shared with some TFs. An overall pictorial representation of the PUF network is provided by Figure 4.

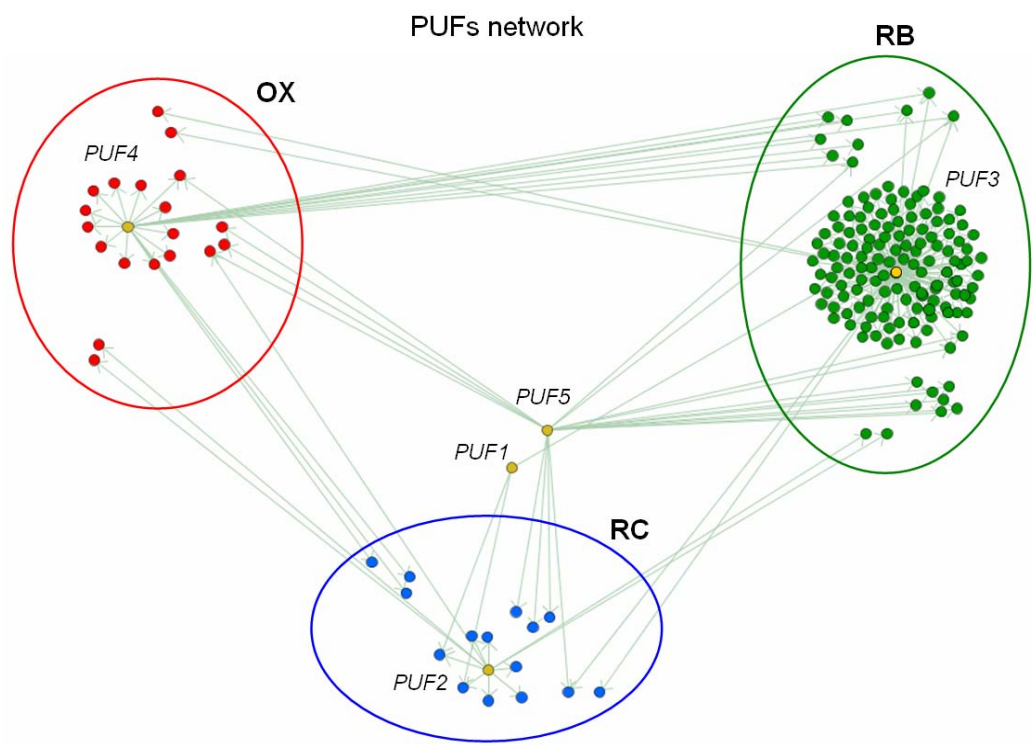

**Figure 4.** Pictorial representation of the PUF network. Colors correspond to different phases of the metabolic cycle: red (OX phase), green (RB phase) and blue (RC phase). Orange circles denote PUF proteins and arrows indicate target genes. The picture makes very clear that the PUF3 gene and the most of PUF3 targets peak at RB phase thus showing the high specificity of the post-trascriptional regulation layer.

In Figure 4 the red, blue and green circles correspond to genes of the different metabolic cycle phases, while the PUF family genes are represented by orange circles. The edges link each PUF with its targets. It is immediate to note the phase specificity of the PUFs together with the positioning of some PUFs in between different metabolic cycle phases so that the possibility for post-trascriptional regulation working in the progression from one phase to another is further made clear. In particular, it is striking that PUF3 results to be the main determinant of PUF family specificity in the regulation of yeast metabolic cycle. The RB phase specificity of PUF3 strongly suggest a role for this regulator in the progression from RB to RC phase. This is consistent with recent findings demostrating that PUF3 acts as a transcript-specific regulatory role of mRNA degradation in yeast. More precisely, PUF3 has been shown to affect stability of COX17 [10] and PET123 [14] transcripts. These considerations together, suggest a role for PUF3 in the rapid down-regulation of its targets at the end of the RB phase and in the up-regulation at the beginning of the same phase which is also consistent with a recent experimental study supporting for a role of PUF3 in the reduction of mitochondrial biogenesis during glucose repression [14] downstream of the TOR signalling pathway [15]. Next, we looked at the TF's that, according to McIsaac *et al.* [13], have PUF genes as targets and we obtained the results reported in Table 1.

| **PUF family member** | **TFs** |
| --- | --- |
| PUF1 (YJR091C) | FKH1, FKH2, NDD1, SWI6, MCM1, UME6 |
| PUF2 (YPR042C) | SFP1, FHL1, RAP1 |
| PUF3 (YLL013C) | ABF1, ROX1, SWI4 |
| PUF4 (YGL014W) | (no data available) |
| PUF5 (YGL178W) | PHO2, MBP1, SWI4, SWI6 |

**Table 1.** Transcription factors binding PUF family gene promoters according to McIsaac [] using a stringent $P$ value for binding of 0.001.

The results regarding PUF3 gene are particularly intriguing: both ABF1 and SWI4 have an expression peak in RB but it seems that transcriptional and post-transcriptional regulation are not simply related given that PUF3 has only 10% of shared targets with ABF1 and virtually no common target with SWI4 and ROX1. Moreover, ABF1 is a very general TF, thus the "two arms" (transcriptional and post-transcriptional) seem to be linked in a very complex way. This appears to be consistent, on the one hand, with the results of Cheadle *et al.* [16] where *alternate* regulation of mRNA transcription and mRNA stability was observed during human jurkat T cell activation. On the other hand, the observed behaviour after a carbon source shift in yeast is *simultaneous* regulation of transcription and degradation since a decrease in transcriptional activity and an increase in messenger stability result in an almost flat mRNA abundance time profile [3].

**Conclusions**

The picture emerging from our integrative analysis of the yeast metabolic cycle is the presence of a very specific, finely wired, post-transcriptional PUF mediated regulation. The interplay between transcriptional and post-transcriptional regulations observed during the metabolic cycle is consistent with the extreme robustness and reproducibility of such self-sustained cycle. The two mechanisms are certainly co-ordinated, but from our computational analysis we could not deduce a simple relationship between the two modes of regulation.

This picture has also an intriguing interpretation in terms of "car driving". In fact, it is well known that high performances can be achieved only by actively operating the throttle and the brake pedal together, where the obvious mechanical counterparts are mRNA synthesis and degradation, respectively. Despite the fact that such driving style is very expensive both in terms of control strategies complexity and resource consumption, it appears that, on specific occasions, the cell may prefer a non-optimal behaviour from an energetic point of view in order to efficiently accomplish other tasks. For example, it has been shown [17] that efficient re-entering into the cell-cycle from a non-proliferative state as terminal differentiation can be obtained by simply removing appropriate CKIs.

Taken together, the results presented in this computational analysis performed using gene expression time series from the yeast metabolic cycle and PUF family proteins binding data, indicate that the regulation of mRNA stability is a widespread, phase-specific and tightly regulated mechanism for the multi-layer control of gene expression. Our analysis further supports the possibility that post-transcriptional regulation surpasses the richness and complexity of transcriptional regulation in many, if not all, physiological and developmental situations [12].